\documentclass[11pt,
prd,aps,amssymb,amsmath  ,tightenlines
]{revtex4}
\usepackage{hyperref}
\usepackage{graphics}
\usepackage[pdftex]{graphicx}

\usepackage{amsfonts}
\usepackage{amssymb}

\begin{document}

\title{ Application  of the Kelly Criterion to Prediction  Markets  
}
\author{ Bernhard K Meister }
\email{bernhard.k.meister@gmail.com}

\date{\today}
\begin{abstract}
\noindent
Betting markets are gaining in popularity. Mean beliefs 
generally differ from prices in prediction markets. Logarithmic utility is employed to study     the risk and
return adjustments to prices. Some consequences are described. A modified payout structure is proposed.  


A simple asset price model based on flipping  biased coins is investigated. It is shown using the Kullback-Leibler divergence  how   the misjudgment of the bias and the miscalculation of the investment fraction  influence  the portfolio growth rate. 

  \end{abstract}
\maketitle

 
\section{Introduction}
\label{sec:1a}
\noindent
Two questions are tackled in the paper.  First, the Kelly criterion is applied to prediction markets\cite{wolfers2004} to explore the difference between expectation values and prices, which unlike in conventional financial markets\footnote{Bond prices, if one assumes positive interest rates, are also bounded below and above, since for zero discounting, the maximal cash flow over the lifetime of a (non-perpetual) bond is bounded above.} are bounded not just below but also above. This alters dynamics and  investment possibilities.  The Kelly criterion  is named after  John L Kelly\cite{kelly}, who applied    logarithmic utility and more broadly information theory\cite{cover}   to derive an optimal portfolio strategy for betting on horse races or other events with side information. The strategy he proposed maximizes   the capital growth rate as well as the associated logarithmic utility. 
  The identification of prices in prediction markets with probabilities is popular\footnote{The sentence, ``[p]rices (odds) on Polymarket represent the current probability of an event occurring", could  be found  on  December $11^{th}$ on the webpage \url{https://learn.polymarket.com/docs/guides/get-started/what-is-polymarket/}.} but  incorrect.  How large can the gap be, and how can the betting payout be adjusted  to reduce this gap? 

 Second, in a finite time setting the simple  asset price model based on a biased coin model is investigated. Implications for  the growth rate due to the misestimation of the bias or the miscalculation of the   fraction invested in the risky asset are studied.


The paper commences with two sections on prediction markets, followed by a section  on misestimation in biased random walk models, rounded off by a conclusion. 
\section{Prediction Markets: Prices vs Probabilities}
\label{sec:1a}
\noindent 
Betting markets are in the news. During the recent American election campaign   Polymarket, hosted on the Solana blockchain,  saw the placing of bets   of several billion   dollars  on   multiple races and various outcomes. Questions were again raised, after being earlier discussed by Manski\cite{  manski} and others, about reliability of prices as  a guide for expected outcomes and more recently especially about manipulability. Feedback mechanisms  between observed prices, donor and voter behaviour was  a subject of speculation. One issue sometimes ignored is that betting prices of simple `0-1' betting markets do not normally match the  probabilities that such   events occur\cite{wolfers2006}\footnote{An informative summary of the argument with examples  is     Quantian's blog from October $24^{th}$: \url{https://quantian.substack.com/p/market-prices-are-not-probabilities } }.
This follows from basic considerations in finance and betting. 

In finance one distinguishes `real probabilities' from `market probabilities' which incorporate the  risk aversion of the mythical marginal investor. This has been extensively discussed in the context of `changes of measure' and the inclusion of a pricing kernel in the calculation  of the expectation value of a financial assets reflects this fact.

In betting,  the asymmetry of the payouts is of importance. A bet away from $50\%$ gives one side a larger return than the other side. This means  leverage and  capital adjusted returns differ, which 
 influences the odds. 

The first argument sees the prediction market as part of a larger family of investable assets shaping the overall preferences of investors. The difference between systematic and idiosyncratic risk comes to mind, whereas the second argument  also applies  to bets in isolation. The second route is chosen for its simplicity in the following section. To wager  on even or odd number sunspots during a time interval is an example of a simple  uncorrelated binary  market bet.


\section{Differences between  prices and probabilities  in prediction markets }
\noindent
In this section  logarithmic utility is applied to obtain the optimal investment fraction, when the market price is $p$, but the subjective  belief of the investor is $q$.  Afterwards we look at  applications. 
Some background about prediction markets. In an `all-or-nothing' contract, if one pays $p$, a value between zero and one dollar, to go long one unit, then one receives one dollar, if the event happens, but loses the initial outlay, if it does not occur at the terminal time. Often  the terminal time is fixed, but sometimes there is a certain flexibility. For the American election bets there were different resolution mechanisms, e.g.  a  subset of national television networks had to declare a candidate a winner for a payout to occur. The initial price $p$  is the maximal loss associated with the contract and kept by the exchange until the contract ends. There is  therefore only an initial but no variation margin. This lack of dynamic margining\footnote{ Dynamic margining in an anonymous setting, preferred by many market participants, is harder to implement, but this hasn't stopped decentralised finance venues from adopting  a variety of  approaches.} reduces the capital efficiency for investors and might require adjustment  in future iterations of prediction markets. The inability to have naked short positions leads in binary outcome markets to the trading of two linked contracts. In the following analysis, the  existence of mirror markets  as well as the existence of fees and other costs is ignored. We also skate over the actual mechanics of the market, which are  tricky, since centralised and decentralised   smart contract controlled elements  hosted on the blockchain are intermingled.  Instead we regard the market as a black box that `works'.

The optimal Kelly fraction for an `all-or-nothing' contract with a prohibition on borrowing follows from finding the extremum of the logarithmic utility function $U$:
\begin{eqnarray}
U( q, p,f)&=&(1-q) \log (1-f ) + q \log \Big(1+f \frac{1-p}{p}  \Big)\nonumber\\
 \frac{\partial U(q,p,f)}{\partial f}&= &q \frac{\frac{1-p}{ p}}{1+f \frac{1-p}{ p}  }  -\frac{1-q}{1-f }   =0 \nonumber\\
\Rightarrow f&= &  q - p  \frac{1-q}{1-p}= \frac{ Q-P} {1+Q}.  \nonumber 
\end{eqnarray}
with $P=\frac{p}{1-p} $ and $Q=\frac{q}{1-q} $.

Next we consider   the maximal achievable gap   between the observed price and the mean belief, which we define as the capital weighted expectation value of all the investors active in the market. Here we assume that all investors are logarithmic utility maximizers, who view the particular prediction market in isolation from other activities, and the contribution of the $i^{th}$   investor to the expectation value, i.e. mean belief,   is their subjective probability $q_i$ times their capital $C_i$. Therefore, if there are $N$ investors the mean belief is
\begin{eqnarray}
\sum_{i=1}^N  C_i q_i,  \nonumber 
\end{eqnarray}
while their contribution to market activity, i.e. their position in the market,  is $C_i f_i(p,q_i)$.
For the market to clear the sum $\sum_i C_i f_i(p,q_i)$ has to be zero at the market price $p$.
This leads to a simple scaling relations, for example increasing the capital of all participants by the same factor will not affect $p$. 
To get a feeling for possible bounds we assume a simplified market of just two investors. One being certain of the outcome and the other having  the belief $q$.
 How big can the gap be in this   case? 
There are two cases to consider. Let us first assume that one investor is certain that the prediction will not come true, and afterwards that one investor is certain the event will   occur.
In this case market clearance is achieved, assuming the supremely confident investor has capital of one dollar, by
\begin{eqnarray}
  &&0=C^-_1 f(q^-_1, p)    -1 = C^-_1  \frac{Q^-_1-P}{1+Q^-_1}  -1 ,  \nonumber \\
  &&\Rightarrow C^-_1 =\frac{Q^-_1+1}{Q^-_1-P}=\frac{1-p}{ q^-_1-p},
\end{eqnarray}
with  $Q^-_1=\frac{q^-_1}{1-q^-_1} $ which leads to a expectation value of 
\begin{eqnarray}
  &&E_- (q^-_1,p)= \frac{C^-_1 q^-_1  }{1+C^-_1} =\frac{ 1-p }{q_1^--2p+1}q_1^- .  \nonumber \\
\end{eqnarray}
 The different possible choices of $q^-_1$ give in this case for $E_-$ a range of outcomes, which are $[ 0 ,\frac{1}{2} ]$ for $p<1/2$,  $[ \frac{1}{2},1 ]$ for $p>1/2$, and $1/2$ for $p=1/2$.
 
 \noindent
 Let us next assume that one investor is certain the event will happen.
The equations for market clearance is
\begin{eqnarray}
  &&0=C^+_1 f(q^+_1, p)  + 1 = C_1  \frac{Q^+_1-P}{1+Q^+_1}  +1 ,  \nonumber \\
  &&\Rightarrow C^+_1 =\frac{Q^+_1+1}{P-Q^+_1}=\frac{1-p}{p-q^+_1}, 
\end{eqnarray}
with  $Q^+_1=\frac{q^+_1}{1-q^+_1} $, which leads to an expectation value of 
\begin{eqnarray}
  &&E_+ (q^+_1,p)= \frac{C^+_1 q^+_1 + 1}{1+C^+_1}=p. 
 \nonumber 
\end{eqnarray}
Even these toy examples show that the gap can be  wide.
Another extreme example, if all investor have either a belief of $0\%$ or $100\%$
in the outcome, then even the minutest imbalance drives the price to zero or one, but the expectation value does not need to budge much from $1/2$.
One can also turn the problem around and fix $q_1$ and vary $p$. The  calculation is similar to the above and will not be carried out.
A more complicated investor structure could also be studied, but it would not change the general insight that prices and beliefs can differ markedly. 





Can one structure a betting market, which 
 is better at attracting liquidity?
Instead of trying to find an optimal model, we just show how a simple change in the payout can change the utility and the optimal fraction:
\begin{eqnarray}
&&U_{\alpha}(q,p,f)= (1-q) \log (1-f ) + q\log \Big(1+f \Big(\frac{p}{1-p}\Big)^\alpha  \Big)\nonumber\\
&& \frac{\partial U_{\alpha}(q,p,f)}{\partial f}=  \frac{ \Big(\frac{p}{1-p}\Big)^\alpha}{1+f \Big(\frac{p}{1-p}\Big)^\alpha }  -\frac{1-q}{1-f } =0 \nonumber\\
&&\Rightarrow  f=\frac{ \hat{P}_{\alpha}-Q}{ Q+1}  \nonumber 
\end{eqnarray}
with $\hat{P}_{\alpha}=\Big(\frac{p}{1-p}\Big)^\alpha$, where for $\alpha=1$ one gets the conventional result, and for $0\leq \alpha\leq 1$ one gets a smoothing and reduction of the gap close to a price of $1$ and the opposite close to $0$, whereas for $\alpha\geq 1$ the behaviour close to prices $1$ and $0$ is reversed. As an example, if $\alpha$ moves from one towards zero the high probability bets with low relative pay-out become more attractive. 
This suggests that if the initial price is close to either boundary one chooses an appropriate $\alpha$ to increase the attractiveness, and as a consequence the liquidity of the market.  These adjustments are not a panacea, since if the market flips to the other extreme liquidity would drain away. 
 The modification of $\alpha$ effects the gap between prices and mean beliefs. Various cases have to be considered and the result will be presented elsewhere.   

\noindent
Next a section on misestimation in biased random walk models.
\section{The relative performance of different strategies in the double-or-nothing game}
\noindent
Various measures, see for example Breiman\cite{breiman}, have been proposed to compare the performance of a portfolio following the Kelly criterion with other strategies.
In the long-term the portfolio growth of the Kelly investor beats all other strategies. What happens over finite times is studied in this section.
The  biased discrete random walk of $N$ steps with a probability $p$ has the following probability mass function
\begin{eqnarray}
 \binom{N}{k}p^k (1-p)^{N-k}
 ,\nonumber 
\end{eqnarray}
where $k$ is the number `upward steps', 
and this allows the definition of  
\begin{eqnarray}
F(k,N,p) =\sum_{i=0}^{\lfloor k\rfloor} \binom{N}{i} p^i (1-p)^{N-i},
 \nonumber 
\end{eqnarray}
which is the probability that number of `upward steps' is less or equal to $k$.
There are various ways to interpolate sums between integers, i.e replace $\lfloor k\rfloor$ by $k$ and still have a meaningful definition of the sum, but the gain would be limited. 
Here instead we just replace $k$ by the nearest smaller natural number and, where possible, choose other parameters such that $k$ is a whole number. 
The Chernoff inequality gives the  bound 
\begin{eqnarray}
F(k,N,p) \leq e^{-N D \big(\frac{k}{N}
\parallel p\big)}
 ,\nonumber 
\end{eqnarray}
with 
\begin{eqnarray}
  D \Big(\frac{k}{N}\parallel p\Big)= \frac{k}{N} \log \Big( \frac{k}{pN }\Big) +  
  \Big( 1-\frac{k}{N}\Big) \log \Big( \frac{1- \frac{k}{N}}{1-p}\Big),\nonumber
\end{eqnarray}
which goes by the name of  Kullback-Leibler divergence function and relative entropy.
Besides an upper bound one can also obtain a lower bound\footnote{The inequalities of this section can be found on the Wikipedia page for binomial distributions.} of the form
  \begin{eqnarray}
F(k,N,p) \geq\frac{1}{\sqrt{2N}} e^{-N D \big(\frac{k}{N}
\parallel p\big)}
.\nonumber 
\end{eqnarray}
Next the logarithms is applied to $F(k,N,p)$ and the result is divided by $N$. For large $N$ one gets  approximately $D( \frac{k}{N}\parallel p)$ for both the lower and upper bound, which   is  linked to the `rate function' from large deviation theory. 
Furthermore we choose  the value $k$ to be  
\begin{eqnarray}
k_Q(f)=  \frac{Q -N \log(1-f)  }{\log (1+f)  -\log(1-f)}.
\nonumber
\end{eqnarray}
due to 
\begin{eqnarray}
(1+f)^{k_Q} (1-f)^{N-k_Q}=  e^{Q}.
\nonumber
\end{eqnarray}
The Kullback-Leibler divergence $D( \frac{k}{N}\parallel p)$ is now a tool to analyze the performance of different strategies in the double-or-nothing game.
 The sensitivity analysis of the expression $D( \frac{k_Q }{N}\parallel p)$ is carried out in two cases, which correspond to the misestimation of the probability and the misjudgment of the optimal investment fraction.
The first case corresponds to  holding $f$  fixed and varying the probability $p$ in $D \Big(\frac{k  }{N}\parallel p\Big)$ and the second one to  holding $p$ fixed and varying the investment fraction $f$.
If $p $  is  varied by $\epsilon$, then one  has 
\begin{eqnarray}
&&D \Big(\frac{k }{N}\parallel p+ \epsilon\Big)- D \Big(\frac{k  }{N}\parallel p\Big)
= \frac{p -\frac{k}{N}}{p(1-p)  }\epsilon+O(\epsilon^2).\nonumber 
   \end{eqnarray}
 If $k$ is varied, then the expression is more involved, but the leading term in the expansion is again  linear. In contrast, if one varies the entropy $U(p,f)$, which corresponds to the portfolio growth rate, around the optimal fraction $2p-1$, then the leading term is quadratic, i.e. 
  \begin{eqnarray}
  U(p,2p-1+\epsilon)- U(p,2p-1) =-\frac{1}{4(1-p)p} \epsilon^2+ O(\epsilon^3).
  \nonumber
  \end{eqnarray}  
Extensions to the quantum case\cite{meister2024} are possible, and will be carried out elsewhere. 
\noindent
Next, there is a brief conclusion to round off the paper. 

\section{Conclusion}
\noindent
Information theory  even in its simplest form can shed light on prediction markets. It was shown how mean beliefs differ from prices, and how the market might be modified by stretching or compressing the pay-out ratio   close to the boundaries.  

The Kullback-Leibler divergence, a tool to analyze the performance of different strategies in the double-or-nothing game, 
 was applied to calculate differences in the achievability of growth thresholds  due to  the misestimation of the random walk bias and the miscalculation of the portfolio fraction.


\begin{enumerate}


\bibitem{breiman}Breiman, L,  ``Optimal gambling systems for favorable games." The Kelly Capital Growth Investment Criterion (1961): 47-60.
\bibitem{cover} Cover, T~M,  \& J~A Thomas, ``Elements of Information Theory."    New York: Wiley-Interscience, (1991).
\bibitem{kelly}Kelly, J~L, ``A New Interpretation of Information Rate." Bell System Technical Journal. 35 (4) (1956): 917–926. 
\bibitem{manski} Manski, C, “Interpreting the Predictions of Prediction Markets.”, NBER
Working Paper \#10359, (2004).
\bibitem{meister2024}
Meister, B~K,  \& H~C~W Price, ``A Quantum Double-or-Nothing Game: An Application of the Kelly Criterion to Spins." Entropy 26.1 (2024): 66.
\bibitem{wolfers2004} Wolfers, J,  \& E  Zitzewitz, ``Prediction markets." Journal of economic perspectives 18.2 (2004): 107-126.

\bibitem{wolfers2006} Wolfers, J,  \& E  Zitzewitz, ``Interpreting prediction market prices as probabilities." (2006).



























































































%














	
\end{enumerate}
\end{document}